\documentclass[prb,twocolumn,showpacs,amsmath,amssymb]{revtex4}

\usepackage{graphicx}
\usepackage{dcolumn}
\usepackage{bm}
\begin{document}
\title{Magnetic groundstate and Fermi surface of bcc Eu}
\author{J. Kune\v{s}}
\email{kunes@fzu.cz}
\affiliation{Dept. of Physics, University of California, One Shields Avenue,
Davis CA 95616, USA \\ and
Institute of Physics, Academy of Sciences of the Czech Republic, Cukrovarnick\'a
10,162 53 Praha 6, Czech Republic}
\author{R. Laskowski}
\affiliation{Institut for Fysik og Astronomi Aarhus Universitet,
Ny Munkegade, 8000 Aarhus C, Denmark}
\date{\today}

\begin{abstract}
Using spin-spiral technique within the full potential linearized augmented-plane-waves
(LAPW) electronic structure method we investigate the magnon spectrum
and N\'eel temperature of bcc Eu. Ground state corresponding to an 
incommensurate spin-spiral is obtained in agreement with experiment and
previous calculations. We demonstrate that the magnetic coupling is primarily
through the intra-atomic $f-s$ and $f-d$ exchange and Ruderman-Kittel-Kasuya-Yosida
mechanism. We show that the existence of this spin-spiral is
closely connected to a nesting feature of the Fermi surface which was not noticed before.
\end{abstract}

\maketitle

\section{Introduction}
The magnetic behavior of most rare-earth (RE) materials is governed
by localized magnetic moments interacting indirectly through
the sea of delocalized valence electrons. A typical feature of RE
systems is existence of two electron species, localized $4f$'s 
exhibiting atomic like behavior with strong Coulomb interaction
and delocalized $5d$ and $6s$ electrons with interaction merely
renormalizing the band dispersion. Yet, a typically weak interaction 
between these two species, either due to intra-atomic exchange or
band mixing (hybridization), gives rise to a variety of magnetic 
behaviors \cite{jen91}. On the model level this behavior is
captured by periodic Anderson model or Kondo lattice model \cite{hew93}.
The {\it ab initio} electronic structure methods 
based on density functional theory (DFT) \cite{hoh64} and the
standard semi-local approximations \cite{koh65,per92} have notorious problems
in dealing with the strong correlations within the $4f$ shell. In particular
the splitting between occupied and unoccupied $4f$ bands, which is the way 
a single-particle band structure can capture the effect of 
Coulomb interaction, is missing. Consequently, the $f$ bands appear
at the Fermi level resulting in unrealistic filling of both $f$ and
valence orbitals. Two remedies can be used: (i) open-core treatment
or (ii) additional Coulomb term with the simplest example being the LDA+U method.
In the open-core treatment the $4f$ orbitals are kept separate
from the rest of the valence Hamiltonian and the interaction
with the valence states is only through the self-consistent potential.
Obviously the proper filling of both $f$ and valence bands is easy
to achieve if integer, however all kinematic exchange effects (e.g. superexchange)
based on band mixing are missing. In the LDA+U approach the splitting
between the occupied and unoccupied $f$ band is obtained due to an
additional orbital-dependent term. All possible exchange processes 
are in principle accounted for in this approach. 

Compounds containing Eu in $2+$ formal valency are particularly
well suited for LDA+U treatment, since the orbital degrees of freedom
are quenched in the half filled $f$ shell. Examples involve 
ferromagnetic insulators EuO and EuS \cite{wac79}, semimetal EuB$_6$ \cite{eub6}
and metallic elemental Eu. Presumably the exchange mechanisms in these materials
are quite different.

In this paper we investigate the magnetic ground state and magnon
spectrum of elemental Eu. Europium crystallizes in body-centered cubic 
(bcc) structure with a lattice constant\cite{ner64} of 4.555 \AA \ at 100 K. 
The magnetic groundstate was found to be a spin spiral and
the N\'eel temperature\cite{ner64,mil73} of 91 K. 
The electronic structure was previously investigated by Freeman and Dimmock
\cite{fre66} and Andersen and Loucks \cite{and68} using the $X\alpha$
potential. Recently Turek {\it et al.} \cite{tur03} used a real-space
perturbation approach based on tight-binding linear muffin-tin orbital
(TB-LMTO) method to calculate the exchange parameters and corresponding
magnon spectrum and N\'eel temperature. Here we use a reciprocal
space based spin spiral approach which can be viewed as complementary
to the real space calculations. Unlike the above authors who used the
open core treatment of the $4f$ orbitals we employ the LDA+U method.
We use the linearized augmented-plane-waves (LAPW) \cite{sin94} method and its
extension to non-collinear magnetic structures utilizing the 
generalized Bloch theorem \cite{san98} for calculation of the spin spiral states.
We interpret our results in terms of Ruderman-Kittel-Kasuya-Yosida (RKKY)
exchange mechanism \cite{rud54,yos57,kas56} and the Fermi surface property. 
For this purpose we
evaluate the low frequency limit of the generalized susceptibility and
identify the nesting features of the Fermi surface.  

\section{Computational method}
We have used the 
Wien2k \cite{wien2k} implementation of the full potential
LAPW method and its extension for non-collinear spin structures \cite{las}.
The effective single-particle potential with constructed from LDA+U functional, with
the exchange-correlation potential of Perdew and Wang \cite{per92} 
and the double-counting scheme of Anisimov {\it et al.} \cite{ani93}.
The Coulomb term parametrized with U (7 eV unless stated otherwise) and
J (0.75 eV) was applied to the $4f$ orbitals. 
The spin-spirals were treated using the generalized
Bloch theorem \cite{san98} which prohibits inclusion of the spin-orbit 
coupling. The atomic sphere approximation was employed, 
in which the direction of the exchange field is constrained inside
atomic spheres and allowed to vary continuously in the interstitial space.
\begin{figure}
\includegraphics[angle=270,width=10cm]{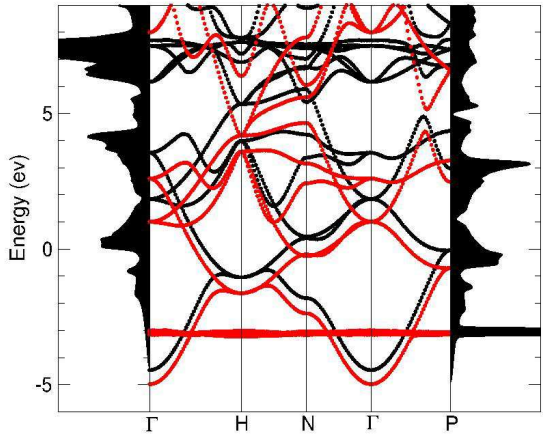}
\caption{\label{fig:band_ldau}(color online) The band structure as obtained with LDA+U method. The
brighter (red) lines correspond to majority spin. In the side panels the corresponding
densities of state for the minority (left) and majority (right) spin projections
are shown.}
\end{figure}

A spin spiral is defined by a propagation vector $\mathbf q$
and angle $\theta$ between the local moment and the precession axis. 
The orientation of the precession axis itself is arbitrary
unless the spin-orbit coupling is taken into account.
Each spin spiral was calculated selfconsistently.
This approach has two deficiencies compared the perturbative approach employing 
the force theorem \cite{lich87}:(i) it costs much more computational effort,
(ii) more importantly one has to work with total energies instead
of the sums of eigenvalues, i.e. looking at small differences of
large numbers which requires high accuracy. As for (i), when performing
calculations for q-vectors along a certain path in the reciprocal 
space converged spin density from a nearby $\mathbf q$ can be used
as starting point, which reduces the number of iterations needs significantly
compared to starting from scratch for each $\mathbf q$.
As for (ii), $\theta$ dependence of spin spiral energies can be studied
without being limited to small angles.
\begin{figure}
\includegraphics[angle=270,width=10cm]{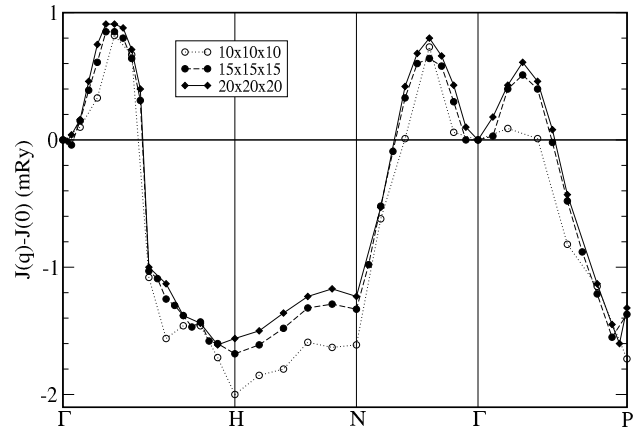}
\caption{\label{fig:magnon}The q-dependent exchange parameter calculated using the spin-spiral
approach. Comparison of the results for different k-point samplings indicates
that the 15x15x15 mesh is reasonably well converged.}
\end{figure}

\section{Results and discussion}
\subsection{Bandstructure}
In Fig. \ref{fig:band_ldau} we show the spin-projected bandstructure 
obtained with U of 7 eV and J of 0.75 eV.
The lowest valence band around the $\Gamma$ point with a 
predominant $6s$ character corresponds to a strongly dispersive
$s$-band. Moving toward the zone boundary mixing with $d$-band takes place
and the doublets at H and P points have a pure $d$ symmetry.
The states in the vicinity of the Fermi level have mostly $d$ character.

The occupied $4f$ levels are localized about 3 eV below the Fermi level and cross
the lowest valence band, with negligible hybridization, which is reflected by
completely flat dispersion. The unoccupied $4f$ bands are approximately 7 eV
above the Fermi level. A bandwidth of about 2 eV originates from mixing
with the $6p$ and $5d$ bands. Lack of mixing
with valence states in the occupied $f$ bands indicates that the kinematic exchange, 
involving hopping from localized $f$ orbitals into the delocalized band states, 
is not important here.
The interaction between the localized $f$ states and the rest of the electronic
system, is dominated by intra-atomic $f-s$ and $f-d$ exchange.
Consequently the spin polarized bands and density of states below 2 eV exhibit
almost a perfect rigid shift. The deviations from this pattern are mostly
due to 
the splitting of the $s$-like bands being less than that of $d$-like bands,
because the more extended $s$ orbital has smaller overlap with
the polarized $f$ density.

\subsection{Magnon spectrum}
The magnetic excitations are discussed in terms of a classical Heisenberg
Hamiltonian
\begin{equation}
\label{eq:hh}
H=-\sum_{\mathbf R\mathbf R'}J_{\mathbf R\mathbf R'}{\mathbf e_{\mathbf R} 
\cdot \mathbf e_{\mathbf R'}}.
\end{equation}
A spin spiral characterized by the propagation vector $\mathbf q$ and angle
$\theta$ has the form
\begin{equation}
\label{eq:the}
\mathbf e_{\mathbf R}=(\sin(\theta)\cos({\mathbf q\cdot \mathbf R}),
\sin(\theta)\sin({\mathbf q\cdot \mathbf R}).
\cos(\theta)),
\end{equation}
The corresponding energy per lattice site obtained from (\ref{eq:hh})
\begin{align}
\label{eq:eq}
E({\mathbf q})&=\sin^2(\theta)\bigl(J({\mathbf q})-J(0)\bigr)+J(0)+E_0 \\
J({\mathbf q})&=\sum_{\mathbf R} J_{\mathbf 0R}\exp(i{\mathbf q\cdot \mathbf R}).
\end{align}
is to be compared to the {\it ab initio} results. $E_0$ is the non-magnetic 
part of the total energy. Note that only the difference $J({\mathbf q})-J(0)$
can be obtained from the knowledge of $E({\mathbf q})$. In order to 
fix the value of $J({\mathbf q})$ the sum rule 
\begin{equation}
\label{eq:sum}
\int d{\mathbf q}J({\mathbf q})=0, 
\end{equation}
where the integration is over the Brillouin zone, is to be employed.
The $J({\mathbf q})$ normalized this way contains only information
about the inter-site $J_{\mathbf R\mathbf R'}$ and can be related to
the ordering temperature, which we discuss below. 
\begin{figure}
\includegraphics[angle=270,width=10cm]{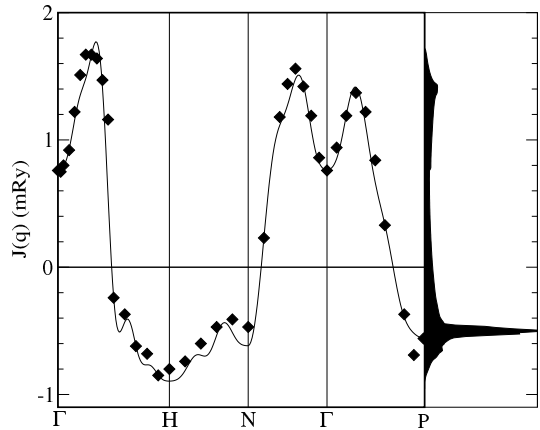}
\caption{\label{fig:jqfit}The q-dependent exchange parameter renormalized to satisfy the sum rule
(\ref{eq:sum}) along the high symmetry lines; {\it ab initio} data (symbols),
Fourier interpolation (line). The right panel shows the corresponding density
of states.}
\end{figure}

Using the Hamiltonian \ref{eq:hh} involves several approximation. Quenching
of the orbital moment in a half filled shell, rigidity of the $f$ moment
and its size (S=7/2) well justify the use of Heisenberg Hamiltonian in classical 
approximation. In addition we assume that the exchange parameters are constant.
This is not {\it a priori} guaranteed, since the electronic structure of the
band electrons, which carry the exchange interaction between the local $f$ moments,
depends on the arrangement of local moments. This question was discussed in detail 
by Nolting {\it et al.} \cite{nol97} who derived an expression for the effective 
exchange parameters $J_{\mathbf R \mathbf R'}$ in terms of the conduction electron self-energy.
Experimentally this leads to temperature dependence of the
effective exchange parameters. If this effect were important a deviation
from the $\theta$ dependence of eq. (\ref{eq:eq}) is expected. The fact
that we have not found any significant deviation from (\ref{eq:eq}) in the
range from $90^o$ to $30^o$ serves as a justification of use of Hamiltonian (\ref{eq:hh}).
This is also agreement with a rather big ratio of the band-width
to the exchange splitting in the conduction band.

In Fig. \ref{fig:magnon} we show the q-dependent exchange parameter
obtained with maximum $\theta$ of $90^o$. 
Calculations performed with U of 6 eV and 8 eV lead to almost identical
dispersion supporting our previous conclusion about the $f-d$ exchange mechanism
which does not depend on the position of the $f$ bands. The calculations
yield a minimum energy corresponding to a spiral with propagation vector
$\mathbf Q$ of about $0.3\times (\frac{2\pi}{a},0,0)$ and another two local
minima along $\Gamma$N and $\Gamma$P lines.
\begin{figure}
\includegraphics[angle=270,width=10cm]{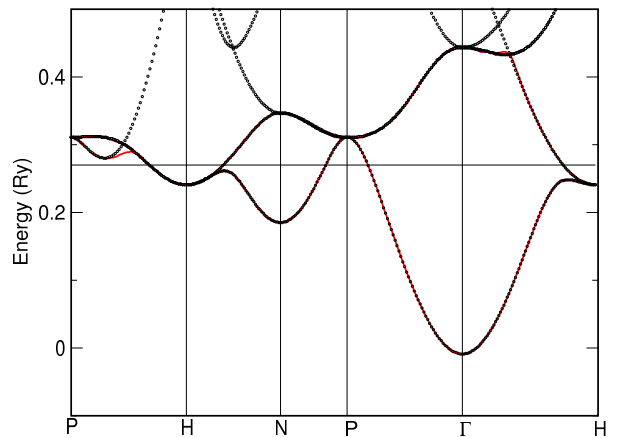}
\caption{\label{fig:band_para}(color online) The paramagnetic {\it ab initio} bands (dots) together
with the smooth Fourier interpolation of the low valence bands (red lines). 
Note that the inaccuracies of the fit at the band crossings 
(for example midway between P and H) are far from the Fermi level.}
\end{figure}
`
In order to address the N\'eel temperature the $E(\mathbf q)$ throughout 
the Brillouin zone is needed. 
To this end we have calculated the spin-spirals on a $10\times10\times10$ regular
q-grid (44 irreducible q-points) and used a smooth Fourier interpolation \cite{pic88,sha71,koe86}
to obtain $J({\mathbf q})$ on a denser grid. In Fig. \ref{fig:jqfit} we compare
the interpolating function to the {\it ab initio} results along the 
high symmetry directions. Since only a few high symmetry q-points were
contained in the regular grid the agreement between the interpolating function
and the {\it ab initio} data points indicates the quality of the grid.
Comparison to the results obtained by Turek {\it et al.} reveals a very good 
agreement of the key features. Besides the similar overall shape of the spectrum,
we have obtained the same ordering of the peak according to their size as well
as similar absolute values. The most apparent deviation is the position
of the ferromagnetic state $J(0)$ which is relatively more favorable in our 
calculation. 
\begin{figure}
\includegraphics[angle=270,width=8cm]{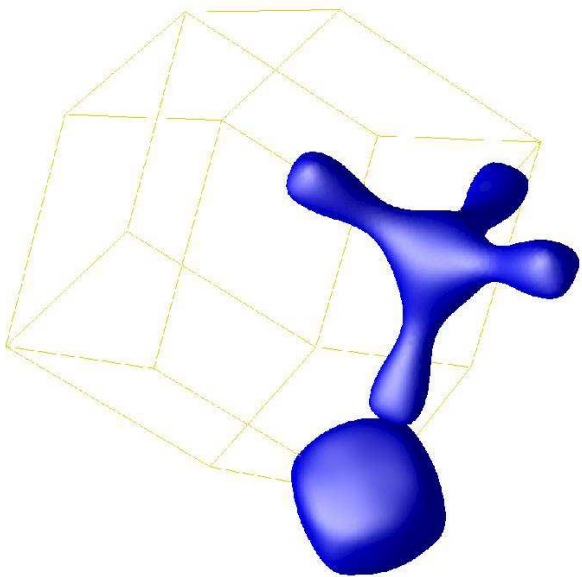}
\caption{\label{fig:fs}The paramagnetic Fermi surface: the lobed tetrahedra are centered at 8 P
points and 'supereggs' at 6 H points on the surface of the 1st Brillouin zone
(only one representative of both species is shown for sake of clarity).}
\end{figure}
Evaluating the expressions \cite{tur03} for N\'eel temperature in the 
mean-field ($T_N^{MF}$) and random-phase ($T_N^{RPA}$) approximations
\begin{gather}
k_BT_N^{MF}=\frac{2}{3}J({\mathbf Q})\\
\begin{split}
(k_BT_N^{RPA})^{-1}&=\frac{3}{4}\frac{1}{N}\sum_{\mathbf q}
\biggl\{\bigl[J({\mathbf Q})-J({\mathbf q})\bigr]^{-1}+ \\
&\bigr[J({\mathbf Q})-\frac{1}{2}J({\mathbf q+\mathbf Q})-
\frac{1}{2}J({\mathbf q-\mathbf Q})\bigr]^{-1}\biggr\}
\end{split}
\end{gather}
with the interpolating function we arrive at $T_N^{MF}$=170 K and
$T_N^{RPA}$=112 K which again agree very well with 147 K and 110 K obtained 
by Turek {\it et at.} as well as with the experimental value\cite{ner64,mil73} of 90.5 K.
\begin{figure}
\includegraphics[width=7cm]{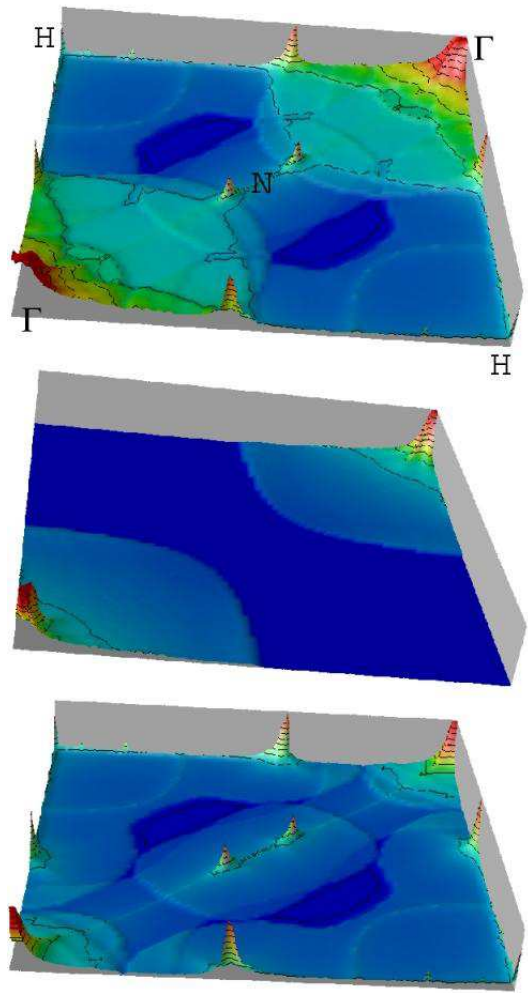}
\caption{\label{fig:ghn}(color online)
The nesting function in the $\Gamma$HNH plane. The lower-left and upper-right
corners correspond to $\Gamma$ point with the typical $1/q$ divergence, the remaining
corners are at H point and the N point is in the center of the square. The upper
panel shows the sum over all (2) bands. The middle panel shows the 'superegg' contribution
and the lower panel the lobed tetrahedra contribution (the contribution of the
superegg-to-lobed tetrahedron processes is not shown).}
\end{figure}

\subsection{Fermi surface and nesting}
In this section we discuss the paramagnetic Fermi surface and its connection to 
the spin-spiral ground state. Very fine k-point sampling is required in order
to study the fine details of the Fermi surface. We adopted the following
approach. Starting with 100 irreducible k-points obtained with LAPW code we have
used the procedure for smooth Fourier interpolation \cite{pic88} for the 
bands at the Fermi energy. We have generated the band energies on a finer mesh 
of approximately $1.5\times 10^6$ k-points in the whole Brillouin zone, which was
used for the calculations reported below. In Fig. \ref{fig:band_para} we show
the paramagnetic bandstructure together with the Fourier interpolation, which
is excellent near $E_F$ (note
that the interpolation was performed on different--regular mesh). 

The paramagnetic Fermi surface of bcc Eu, Fig. \ref{fig:fs}, consists of an 
electron pocket centered at H point,
and a hole pocket located at P point. Symmetry related degeneracy of the valence band along
the P-H direction results in touching of these pockets. The first numerical 
investigation of the Fermi surface of bcc Eu was reported by Andersen and Loucks \cite{and68},
who called these pockets 'superegg' and 'tetracube' respectively.
While our calculation reproduces the 'superegg' shape, contrary to Andersen's
cube with lobes we find rather a rounded tetrahedron with lobes.

In order to make connection between the Fermi surface geometry and
calculated spin spiral dispersion we have evaluated the imaginary part
of the generalized susceptibility 
\begin{equation}
\chi_{ij}({\mathbf q},\omega)=-\sum_{\mathbf k}\frac{f(\epsilon_i({\mathbf k}))-
f(\epsilon_j({\mathbf k+\mathbf q}))}{\epsilon_i({\mathbf k})-\epsilon_j({\mathbf k+\mathbf q})-
\omega+i0^+},
\end{equation}
where $\epsilon(\mathbf k)$ is the band energy, $f(\epsilon)$ is the Fermi-Dirac
function and $i$ and $j$ are band indices.
In the limit $\omega\rightarrow 0$ the imaginary part of $\chi({\mathbf q},\omega)$
behaves as
\begin{align}
\label{eq:lim}
\operatorname{Im}\chi_{ij}({\mathbf q},\omega)&=\pi\omega \sum_{\mathbf k}\delta(\epsilon_i({\mathbf k})-
\epsilon_F)\delta(\epsilon_j({\mathbf k+\mathbf q})-\epsilon_F)\\
&=\pi\omega\nu_{ij}({\mathbf q}),
\end{align}
where $\nu({\mathbf q})$ measures so called nesting, i.e. the extent to which 
different parts of the Fermi surface weighted by the inverse square of
the Fermi velocity are parallel. While the real part of the
generalized susceptibility is directly related to the exchange parameters $J({\mathbf q})$
it is the imaginary part which has a straightforward geometrical interpretation.
The real and imaginary parts are bound by the Kramers-Kronig relations, which for
$\omega=0$ limit read
\begin{equation}
\operatorname{Re}\chi(\mathbf q,\omega=0)=\frac{1}{\pi}\int_{-\infty}^{+\infty}
d\omega'\frac{\operatorname{Im}\chi(\mathbf q,\omega')}{\omega'}.
\end{equation}
Comparing this to (\ref{eq:lim}) one can readily see that a peak in $\nu({\mathbf q})$
provides a significant contribution to a peak in $\operatorname{Re}\chi(\mathbf q,\omega=0)$.
In the special case $\mathbf q \rightarrow 0$ limit $\nu({\mathbf q})$ diverges as $1/|q|$, however
the limit $\lim_{q\rightarrow 0}\lim_{\omega\rightarrow 0}\operatorname{Re}\chi(\mathbf q,\omega)$
remains finite, equal to the density of states (or corresponding partial density of states)
at the Fermi level $N(\epsilon_F)$.
In general
peaks in $\nu({\mathbf q})$ indicate a tendency toward instability of the Fermi surface toward
formation of incommensurate structures such as spin or charge density waves. 
\begin{figure}
\includegraphics[width=7cm]{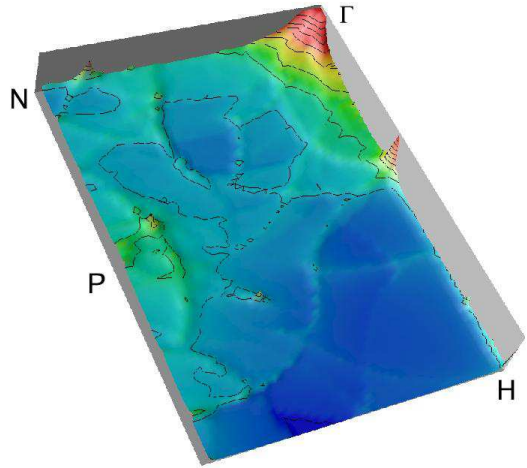}
\caption{\label{fig:ghpn}(color online)
The nesting function in the $\Gamma$HPN plane. The arc shaped features intersecting the
$\Gamma$P line originate from superegg-to-lobed tetrahedron processes.}
\end{figure}
 
\begin{figure}
\includegraphics[angle=270,width=8cm]{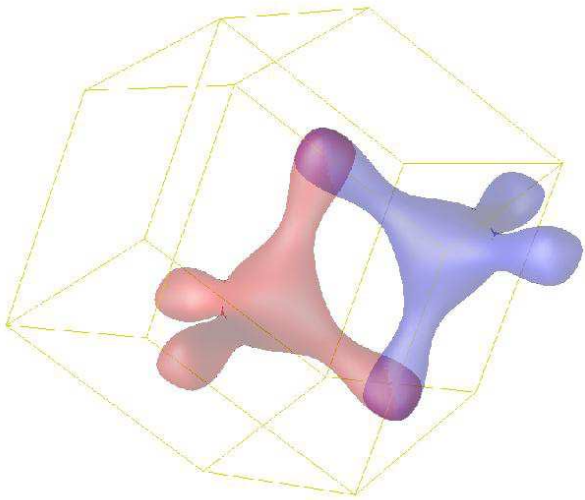}
\caption{\label{fig:nest}The lobed tetrahedron translated by $\mathbf q=0.4\times(\tfrac{2\pi}{a},0,0)$
corresponding to the $\Gamma$-H
peak elucidates the origin of the spin-spiral groundstate.}
\end{figure}

Andersen and Loucks concluded that the origin of the spin spiral groundstate
is the nesting between the opposite faces of their 'tetracube'. However, our investigation
provides a different picture. In Fig. \ref{fig:ghn} we show the nesting
function $\nu({\mathbf q})$ in the $\Gamma$HNH plane. There is a prominent
peak on the $\Gamma$-H line and a weaker feature on the $\Gamma$-N line.
In order to identify the origin of these features we have calculated separately
the contributions of the different sheets of the Fermi surface. Fig. \ref{fig:ghn}
demonstrates that both peaks
originate from transitions between the lobed tetrahedra surfaces.
The $\mathbf q$ parallel to
the $\Gamma$-H direction amounts to moving the lobed tetrahedron centered at the P point
toward the next P point on the same face. The peak on the 
$\Gamma$-H line correspond to the overlap of the lobes,
which is illustrated explicitly in Fig. \ref{fig:nest}.  
The peak on the $\Gamma$-N line corresponds to the sum of two
nesting vectors along adjacent $\Gamma$-H directions. Therefore it 
originates from an overlap of the tips of the lobes,
however, belonging to tetrahedra centered at P  points which are not on the same face. 
In Fig. \ref{fig:ghpn} we show the nesting function in the
$\Gamma$HPN plane. Besides the peaks on $\Gamma$-H and $\Gamma$-N lines we find
weaker features on the $\Gamma$-P line. Analysis of the contributions from
different sheets shows that these features originate from transitions between the
two types of Fermi surfaces.

Eventually we can compare the information obtained from the nesting function
to the magnon dispersion. We have found a prominent nesting feature on the 
$\Gamma$-H line and weaker one on the $\Gamma$-N line. These are related
to the minima in the spin-spiral energies. The energy minima do not sit exactly
at the positions of corresponding the nesting vectors.
While the nesting vectors
contain information about the states directly at the Fermi energy, all states
contribute to the spin-spiral minima, yet with a weight decreasing with the
distance from the Fermi level. 
A discussion of the
generalized susceptibility in the context of electron-phonon coupling can
be found in Ref. \onlinecite{sin75}.The features on the $\Gamma$-P line are
rather weak and do not fit well with the position of the 
$\Gamma$-P peak therefore we do not draw any conclusion
about its direct relation to the local energy minimum on the $\Gamma$-P line.

\section{Conclusions}
Using a full-potential method with LDA+U functional and spin-spiral approach
we have obtained a magnon spectrum which is in good agreement with that
published by Turek {\it et al.} using TB-LMTO. The calculations reproduce
well the experimentally observed spin-spiral groundstate and provide a 
reasonable estimate of the N\'eel temperature. The good agreement of two 
rather different computational methods indicates a robustness of these
physical properties.
Moreover we have shown
that the values of the exchange parameters are insensitive to the precession
angle of the spin spiral as well as the value of U, i.e. exact position of the occupied 
$f$ bands. This confirms the picture of bcc Eu as a Kondo-lattice system
with ferromagnetic exchange of intra-atomic origin in the RKKY regime.
We have identified the origin of the spin-spiral ground state in terms of nesting
properties of the Fermi surface. In particular we have shown that the nesting is 
connected to the lobed tetrahedron hole pockets centered at the 
P point of the Brillouin zone.  

\section{Acknowledgment}
We wish to acknowledge the
helpful discussions and numerous suggestions by W. E. Pickett, J. Kudrnovsk\'y
and P. Nov\'ak. This work was supported by Department of Energy Grant
DE-FG 03-01ER45876, the Grant No. A1010214 from Academy of
Sciences of the Czech Republic and by Czech-USA project KONTAKT ME547.



\begin{thebibliography}{26}
\bibitem[{eub()}]{eub6}
\bibinfo{note}{The questions whether steciometric EuB$_6$ is actually semimetal
  or semiconductor is still controversial}.
\end{thebibliography}
\end{document}